\documentclass[cameraready]{Interspeech}

\title{VOSSA: Voiceprint Optimization for Streaming Speech Architectures}

\author[]{Mu-Ruei}{Tseng}
\author[]{Waris}{Quamer}
\author[]{Ghady}{Nasrallah}
\author[]{Ricardo}{Gutierrez-Osuna}

\address{
    Department of Computer Science \& Engineering, Texas A\&M University, College Station, US
}

\email{\{mtseng,quamer.waris,ghadynasrallah,rgutier\}@tamu.edu}

\keywords{voice conversion, speaker embedding, vowel formants, acoustic analysis}

\usepackage{comment}
\usepackage{multirow}
\usepackage{xcolor}
\usepackage{tipa}
\usepackage{tikz}
\usetikzlibrary{positioning, arrows.meta}
\usetikzlibrary{positioning, arrows.meta, calc}
\usepackage{color, soul}

\begin{document}

\maketitle

\begin{abstract}
Real-time voice conversion (VC) systems commonly rely on pretrained speaker embeddings from automatic speaker verification (ASV) models. While effective for speaker discrimination, these embeddings are trained to remain stable across phonetic and prosodic variations within-speaker, which may conflict with frame-level acoustic generation in streaming constraints. To address this issue, we propose VOSSA (Voiceprint Optimization for Streaming Speech Architectures), a speaker representation framework that extracts speaker information from intermediate content encoder layers and aggregates using attentive statistics pooling. The embedding is trained jointly with VC objectives, removing the need for a separate speaker encoder. Across six datasets, VOSSA improves F0 dynamics and vowel-discriminative acoustic cues while maintaining comparable NISQA-MOS, WER, and speaker similarity. Perceptual tests further indicate improvements in naturalness, speaker similarity, intelligibility, and vibrancy.
\end{abstract}

\begin{figure*}[!t]
  \centering
  \includegraphics[width=\linewidth]{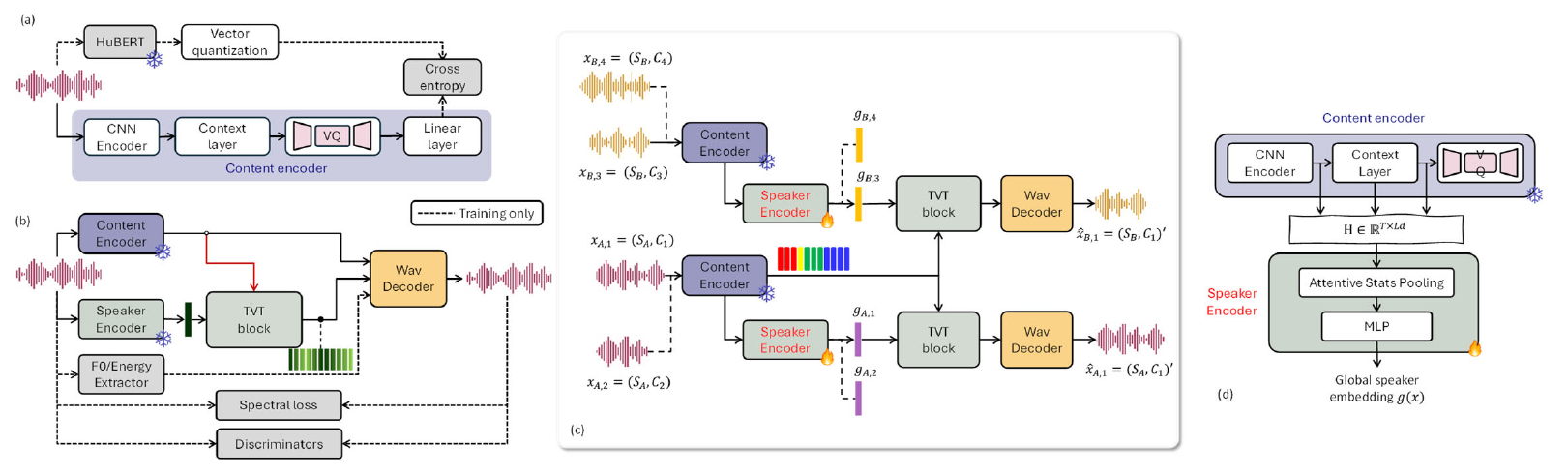}
  \caption{Training workflow for the \ul{TVTSyn backbone}. \textbf{(a)} content encoder trained against  HuBERT k-means pseudo-labels, and \textbf{(b)} decoder conditioned on speaker embedding trained with self-supervision and discriminator objectives. \textbf{(c)} Overview of the \ul{training protocol in VOSSA}. The self-reconstruction path (bottom, purple) uses segments from the same LibriTTS speaker to provide fully supervised training, while the non-parallel VC path (top, orange) uses VoxCeleb targets to condition conversion. \textbf{(d)} Speaker embedding extraction from the frozen content encoder. We collect features from the last CNN layer and every other layer of the MHSA stack, concatenate them to form $\mathbf{H} \in \mathbb{R}^{T \times Ld}$, and apply attentive statistics pooling and an MLP to obtain a global speaker embedding.}
  \label{fig:tvtsyn}
\end{figure*}

\section{Introduction}
Real-time voice conversion (VC) aims to transform speech from a source speaker to match the identity of a known target speaker while generating natural and intelligible speech under strict streaming and latency constraints. Recent streaming architectures achieve latencies as low as 80 ms by combining lightweight content encoders with causal decoders while explicitly disentangling linguistic and speaker representations~\cite{quamer2024end, quamer2025darkstreamrealtimespeechanonymization, quamer2026tvtsyn}. As streaming VC moves toward practical applications, e.g., speech anonymization, VoIP, the design of speaker representations becomes increasingly critical.

VC systems typically decompose a speech signal into linguistic content and a speaker-identity representation, then re-synthesize the content conditioned on a desired target identity. Accordingly, a central design choice is how speaker identity is encoded and injected into the generative model. Existing methods generally follow two approaches. The conventional approach is to condition the decoder on pretrained speaker verification embeddings, e.g., x-vectors~\cite{8461375}, ECAPA-TDNN~\cite{desplanques2020ecapa}, which remain frozen during VC training. More recent approaches learn speaker or style encoders jointly within the generative framework. For instance, GenVC~\cite{cai2025genvc} extracts a fixed-length style embedding through a Perceiver encoder to guide autoregressive generation. Conan~\cite{zhang2025conan} and RT-VC~\cite{liu2025rt} adopt streaming architectures that pair content extraction with adaptive style or speaker encoders for low-latency zero-shot VC. 


Despite their effectiveness, both paradigms present limitations in streaming VC. First, ASV-based speaker embeddings are explicitly trained to suppress intra-speaker variability (phonetic, prosodic), but generative speech modeling requires speaker conditioning to be consistent with frame-level phonetic realizations, such as formant structure. This introduces a representational mismatch between (intra-speaker) invariant identity modeling and phoneme-conditioned acoustic generation. 
On the other hand, approaches that jointly train speaker encoders with the VC model in a self-supervised fashion can reduce this mismatch but they still retain the additional encoder branch for speaker representation learning, increasing model complexity and memory footprint which could be critical in resource-constrained settings. A potential solution to these issues comes from the analysis of large self-supervised speech models such as WavLM~\cite{hsu2021hubert, Chen_2022}, which suggests that intermediate encoder layers retain substantial speaker and acoustic information, while deeper layers become increasingly content-focused. This finding suggests a third, unified, alternative: rather than introducing an additional speaker encoder, speaker identity can be extracted directly from intermediate representations of the content encoder.


To explore this possibility, we present VOSSA (Voiceprint Optimization for Streaming Speech Architectures), an integrated model for speaker representation learning that is suitable for real-time streaming VC. VOSSA derives speaker embeddings from intermediate layers of the content encoder and optimizes them jointly within a generative framework via reconstruction and cross-speaker conversion objectives.
By leveraging representations already computed for content modeling, our approach eliminates the need for a separate speaker encoder while preserving strict causality and low latency. 

We evaluate VOSSA against four streaming VC baselines: three systems that rely on frozen ASV-based speaker embeddings~\cite{quamer2024end, quamer2025darkstreamrealtimespeechanonymization, quamer2026tvtsyn}, and a jointly trained speaker representation model (GenVC~\cite{cai2025genvc}) that learns speaker embeddings within the generative framework. Our formulation achieves performance comparable to state-of-the-art streaming models in terms of NISQA-MOS~\cite{mittag2021nisqa}, word error rate (WER), and harmonicity, while improving target-speaker similarity under normalized embedding evaluation. 
Moreover, to examine the representational characteristics of our model, we perform additional diagnostics based on large-scale spectral statistics, pitch and harmonics analysis, and formant distributions. Across six datasets, the learned representations in VOSSA exhibit richer pitch variability and better preservation of speaker- and vowel-dependent formant distributions, indicating increased phoneme-conditioned acoustic flexibility that is not captured by standard objective scores.


Audio samples are available at our demo page.\footnote{\url{https://morris88826.github.io/VOSSA/}}

\begin{table*}[t]
\centering
\small
\setlength{\tabcolsep}{4pt}
\caption{Evaluation of VOSSA against SOTA streaming VC baselines. 
$\mathrm{Sim}_{\text{src}}^{\text{syn}}$ can be interpreted as anonymization strength (lower is better), whereas $\mathrm{Sim}_{\text{trg}}^{\text{syn}}$ represents VC strength (higher is better). Best results are \textbf{bolded}; second-best are \underline{underlined}.}
\begin{tabular}{lccccc|c}
\toprule
\textbf{Models} & \textbf{Ground truth} &\textbf{slt24}~\cite{quamer2024end} & \textbf{DarkStream}~\cite{quamer2025darkstreamrealtimespeechanonymization} & \textbf{GenVC-s}~\cite{cai2025genvc} & \textbf{TVTSyn}~\cite{quamer2026tvtsyn} & \textbf{VOSSA (Ours)} \\
\midrule
\multicolumn{7}{c}{\textit{Standard VC metrics and harmonics}} \\
\midrule
NISQA-MOS $(\uparrow)$ & $4.06\pm0.84$ & $3.46\pm0.80$ & $3.12\pm0.81$ & $3.04\pm0.83$ & $\mathbf{3.52\pm0.81}$  & \underline{$3.48 \pm 0.91$} \\
WER $(\downarrow)$ & $0.07\pm0.16$ & $0.18\pm0.25$ & $0.25\pm0.27$ & $0.20\pm0.23$ & \underline{$0.17\pm0.23$}  & $\mathbf{0.17 \pm 0.23}$ \\
$\mathrm{Sim}_{\text{src}}^{\text{syn}}$ $(\downarrow)$ & -- & $0.21\pm0.11$ & $\mathbf{0.06\pm0.11}$ & -- & \underline{$0.10\pm0.10$} & $0.11\pm0.31$ \\
$\mathrm{Sim}_{\text{trg}}^{\text{syn}}$ $(\uparrow)$ & -- & $0.46\pm0.11$ & $0.54\pm0.12$ & -- & \underline{$0.59\pm0.11$} & $\mathbf{0.86\pm0.19}$ \\
HNR $(\uparrow)$ & -- & $9.52\pm2.90$ & $7.93\pm3.69$ & $8.67\pm3.04$ & $\mathbf{9.90\pm2.75}$ & \underline{$9.72\pm2.80$} \\
\addlinespace[4pt]
\midrule
\multicolumn{7}{c}{\textit{Pitch prediction \hl{}}} \\
\midrule
Pitch MAE (W) $(\downarrow)$ & -- & $31.4\pm30.4$ & $33.3\pm22.5$ & $30.8\pm27.0$ & \underline{$29.0\pm20.5$} & $\mathbf{27.8\pm21.5}$ \\
Voicing Mismatch (W) $(\downarrow)$ & -- & \underline{$0.25\pm0.10$} & $0.27\pm0.10$ & $0.43\pm0.10$ & $\mathbf{0.23\pm0.09}$ & $0.26\pm0.09$ \\
Pearson's CC $(\uparrow)$ & -- & $0.40\pm0.33$ & $0.21\pm0.30$ & $0.33\pm0.39$ & \underline{$0.40\pm0.31$} & $\mathbf{0.43\pm0.33}$ \\
\addlinespace[4pt]
\midrule
\multicolumn{7}{c}{\textit{Wasserstein distance for vowels (F1), synthesized vs. target}} \\
\midrule
High Vowels $(\downarrow)$ & -- & $42.3\pm31.2$ & \underline{$28.9\pm16.7$} & $32.7\pm20.8$ & $29.2\pm21.7$ & $\mathbf{24.6\pm13.8}$  \\
Mid Vowels $(\downarrow)$ & -- & $44.5\pm29.5$  & \underline{$31.9\pm15.5$} & $49.2\pm25.3$ & $32.9\pm21.8$ & $\mathbf{29.7\pm16.8}$   \\
Low Vowels $(\downarrow)$ & -- & $57.1\pm32.0$ & $37.4\pm24.3$ & $60.7\pm28.2$ & \underline{$36.6\pm21.5$} & $\mathbf{31.0\pm24.5}$  \\
\addlinespace[4pt]
\midrule
\multicolumn{7}{c}{\textit{Mean opinion score (from perceptual listening tests)}} \\
\midrule
MOS $(\uparrow)$ & $4.65\pm0.69$ &  $3.62\pm1.02$ & $2.51\pm1.14$ & $3.16\pm1.18$ & \underline{$3.65\pm0.91$} & $\mathbf{3.79\pm1.01}$\\
\bottomrule
\end{tabular}
\label{tab:main_results}
\end{table*}

\section{Methods}
Our proposed model (VOSSA) uses TVTSyn as a backbone ~\cite{quamer2026tvtsyn}. TVTSyn is a streaming speech synthesizer designed for low-latency VC and anonymization --see Fig.~\ref{fig:tvtsyn}a-b. Its content encoder uses a causal CNN followed by transformer layers with a 2\,s look-back window and 80\,ms look-ahead, producing 50\,Hz frame embeddings quantized through a factorized VQ bottleneck and trained with cross-entropy against HuBERT $k$-means pseudo-labels ($N{=}200$). 
The speaker embeddings generated through external speaker encoders are expanded into a global timbre memory that serves as key and value pairs for input content features to attend to, generating a time-varying speaker embedding representation, which is then passed to the decoder along with the content features.
The decoder mirrors this architecture of the content encoder and reconstructs waveforms via causal transposed convolutions. TVTSyn generates time-varying timbre (TVT) embeddings based on a global speaker embedding conditioned on content tokens. TVTSyn achieves latencies as low as 80\,ms on GPUs. For full architectural details, please refer to~\cite{quamer2026tvtsyn}. VOSSA adopts these components and reformulates how speaker identity is extracted during training as detailed below.
\vspace{-5pt}
\subsection{Joint speaker representation learning}
VOSSA replaces the external speaker embeddings used in TVTSyn (concatenated x-vectors and ECAPA-TDNN) with a speaker representation derived from the same encoder backbone used for content extraction. Let $E_c$ denote the \textit{frozen} streaming content encoder in TVTSyn. Given waveform $x$, we extract content $c = E_c(x)$ and multi-layer speaker features from the last CNN layer and every other layer of the 8-layer MHSA stack (see Figure ~\ref{fig:tvtsyn}d). At each time step $t$, we stack the selected layer features as $\mathbf{H}_t \in \mathbb{R}^{Ld}$, where $L$ is the number of selected layers and $d$ the feature dimension per layer. The full sequence is
$\mathbf{H} = \{\mathbf{H}_t\}_{t=1}^{T}
\in \mathbb{R}^{T \times Ld}$.
We aggregate the resulting feature sequence $\mathbf{H}$ via attentive statistics pooling (ASP)~\cite{okabe2018attentive}. Specifically, we compute attention weights as $\alpha_t = \mathrm{softmax}\big(f(\mathbf{H}_t)\big)$, where $f(\cdot)$ denotes a lightweight fully-connected (FC) two-layer network. Finally, we compute the attention-weighted mean and variance:
\begin{equation}
\mu = \frac{1}{T} \sum_{t=1}^{T} \alpha_t \mathbf{H}_t,
\qquad
\sigma^2 = \frac{1}{T} \sum_{t=1}^{T} \alpha_t (\mathbf{H}_t - \mu)^2.
\end{equation}
and compute the global speaker embedding as $g(x) = \phi([\mu ; \sigma])$, where $\phi(\cdot)$ is a two-layer projection network. This design encourages the embedding to capture utterance-level speaker characteristics while allowing frame-level variability to be weighted adaptively.
\vspace{-5pt}
\subsection{Dual-path speaker-aligned training}
The training protocol for VOSSA is illustrated in Fig.~\ref{fig:tvtsyn}c.  Let us denote an utterance from speaker $S_n$ with content $C_i$ as $x_{n,i} = (S_n,C_i)$, the corresponding speaker embedding as $g(x_{n,i})=g_{n,i}$, and the reconstructed utterance as $\hat x_{n,i} = (S_n,C_i)'$. During training, we randomly select two utterances from each of two random speakers $A$ and $B$: $x_{A,1}, x_{A,2}$ and $x_{B,1}, x_{B,2}$.  The training protocol proceeds in two parallel paths: (1) a self-reconstruction path on the lower branch that reconstructs utterance $x_{A,1}$ at the output, i.e., $(S_A,C_1)'$, and (2) a voice-conversion path that performs voice conversion to speaker $S_B$ using the content from $x_{A,1}$, i.e., $(S_B,C_1)'$.

We extract content features $c = E_c(x)$ and synthesize both self-reconstruction and voice conversion utterances as $\hat{x}_{A} = D(c, \bar{g}_{A})$ and
$\hat{x}_{B} = D(c, \bar{g}_{B})$
where $D()$ denotes the decoder of TVTSyn 
, and $\bar{g}_A$, $\bar{g}_B$ are the average speaker embedding for the two speakers (to promote stable identity targets):
\begin{equation}
\bar{g}_A = \frac{g_{A,1} + g_{A,2}}{2},
\quad
\bar{g}_B = \frac{g_{B,1} + g_{B,2}}{2},
\end{equation}
To enforce consistency within-speakers after re-synthesis, we add a loss term that minimizes the cosine distance $d_{\text{cos}}(\mathbf{u}, \mathbf{v}) = 1 - \cos(\mathbf{u}, \mathbf{v})$ between the original speaker embeddings $(\bar{g}_A, \bar{g}_B)$ and those obtained from the reconstructed utterances:
\begin{equation}
\mathcal{L}_{\text{spk}}
=
d_{\text{cos}}\big(g(\hat{x}_A), \bar{g}_A\big)
+
d_{\text{cos}}\big(g(\hat{x}_B), \bar{g}_B\big).
\end{equation}
As an auxiliary regularization term, we adopt a symmetric NT-Xent (InfoNCE)~\cite{chen2020simpleframeworkcontrastivelearning} objective to encourage embeddings from the same speaker to remain consistent across different utterances. Namely, for each mini-batch, we construct two views by concatenating the source and target speaker representations, yielding $\mathbf{z}_1 = [\mathbf{g}_A^{(1)}; \mathbf{g}_B^{(1)}]$ and $\mathbf{z}_2 = [\mathbf{g}_A^{(2)}; \mathbf{g}_B^{(2)}]$, with corresponding speaker labels $\mathbf{y} = [y_A; y_B]$.
We minimize a bidirectional InfoNCE loss b/w $\mathbf{z}_1$ and $\mathbf{z}_2$. To prevent false negatives, samples sharing the same speaker label are excluded from the negative set, except for the matched positive pair.

The overall training objective is
\begin{align}
\mathcal{L}_{\text{total}}
=
\lambda_{\text{mel}}\mathcal{L}_{\text{mel}}
&+
\lambda_{\text{adv}}\mathcal{L}_{\text{adv}}
+
\lambda_{\text{fm}}\mathcal{L}_{\text{fm}} \notag \\
&+ 
\lambda_{\text{spk}}\mathcal{L}_{\text{spk}}
+
\lambda_{\text{con}}\mathcal{L}_{\text{con}}
\end{align}
where $\mathcal{L}_{\text{mel}}$ is the L1 reconstruction loss, $\mathcal{L}_{\text{adv}}$ and $\mathcal{L}_{\text{fm}}$ are the adversarial and feature-matching  loss from the discriminator, respectively~\cite{quamer2026tvtsyn}, and $\mathcal{L}_{\text{spk}}$ and $\mathcal{L}_{\text{con}}$ enforce speaker consistency and supervised contrastive regularization. Gradients propagate through the speaker encoder, the TVT  module, and the decoder, aligning the learned speaker space with streaming VC.

\section{Experiments}
\label{sec:experiments}

\subsection{Voice conversion protocol and diagnostics}
\label{sec:vc_quality}
To evaluate the effectiveness of the speaker representation learning in VOSSA, we employ two complementary protocols: voice conversion and self-reconstruction.\footnote{The content encoder is kept frozen while the rest were trained on LibriTTS~\cite{zen2019libritts} and VoxCeleb~\cite{nagrani2017voxceleb, chung2018voxceleb2} corpora. We set hyperparameters  $(\lambda_{\text{mel}},\lambda_{\text{adv}},\lambda_{\text{fm}},\lambda_{\text{spk}},\lambda_{\text{con}})=(20,3,1,10,1)$. All modules were trained using AdamW with an initial learning rate of \(1 \times 10^{-4}\), an \texttt{ExponentialLR} scheduler with decay factor \(\gamma = 0.999996\), and a batch size of 64 distributed on four NVIDIA RTX 5000 Ada GPUs.}
For the VC protocol, source utterances are sampled from the test and dev set of LibriTTS~\cite{zen2019libritts}. Target speakers are drawn from LibriTTS, VoxCeleb~\cite{nagrani2017voxceleb, chung2018voxceleb2}, EMIME~\cite{wester2010emime}, ARCTIC~\cite{kominek2003cmu}, L2-ARCTIC~\cite{zhao2018l2}, and VCTK~\cite{veaux2017cstr}. For each target speaker $t$, we sample two disjoint subsets of utterances: a reference set $\mathcal{S}^{ref}_t$ and an evaluation subset $\mathcal{S}^{eval}_t$. Reference utterances provide the target speaker identity for VC, while evaluation utterances are used to extract the speaker’s natural formant distributions. All systems are evaluated on identical source–target pairs.
We evaluate VC performance using conventional objective metrics: speaker similarity, content preservation, and acoustic quality. We measure speaker similarity as the cosine between speaker embeddings of synthesized utterances and both (1) source utterances $\cos(\mathbf{e}_{syn},\mathbf{e}_{src})$ and (2) target utterances $\cos(\mathbf{e}_{syn},\mathbf{e}_{tgt})$. Because the range of cosine similarities for each system is different (e.g., TVTSyn uses pretrained x-vector/ECAPA embeddings while VOSSA learns speaker representations internally), direct comparison of raw cosine similarity is difficult. Instead, we report normalized speaker similarity:
\begin{equation}
\mathrm{Sim}_{\text{trg}}^{\text{syn}} =
\frac{\cos(\mathbf{e}_{syn},\mathbf{e}_{tgt})  - \cos(\mathbf{e}_{src},\mathbf{e}_{tgt})}
{1 - \cos(\mathbf{e}_{src},\mathbf{e}_{tgt})},
\end{equation}
Thus, $\mathrm{Sim}_{\text{trg}}^{\text{syn}}$ represents how close the timbre of synthesized utterance is to that of target speaker \textit{relative} to the timbre similarity between the source and target speakers. Following conventions, we also report word error rate (WER) from a pretrained ASR model~\cite{radford2022whisper}, and acoustic quality using NISQA-MOS~\cite{mittag2021nisqa}.

Additionally, we evaluate voice quality using the harmonics-to-noise ratio (HNR), which is indicative of clearer and more stable vocal production, using Praat's cross-correlation harmonicity method~\cite{boersma2011praat, parselmouth}. We compute frame-level harmonicity values with a 10 ms time step and report the mean HNR (in dB) for each utterance.

Results are summarized in Table~\ref{tab:main_results}. VOSSA achieves comparable NISQA-MOS to TVTSyn, which conditions on frozen speaker embeddings, and outperforms all other baselines, including GenVC-s, which jointly learns speaker representations through a separate encoder branch. VOSSA also matches TVTSyn in WER, confirming preserved intelligibility. In addition, VOSSA achieves higher HNR than most baselines and remains close to TVTSyn, indicating stable harmonic structure and reduced noise in the synthesized speech. Notably, VOSSA achieves significantly higher normalized target similarity, indicating stronger speaker identity transfer relative to the source–target similarity. These findings indicate that the intermediate content representations provide sufficient speaker-discriminative information, eliminating the need for an additional speaker encoder during inference.




\begin{figure}
    \centering
    \includegraphics[width=\linewidth]{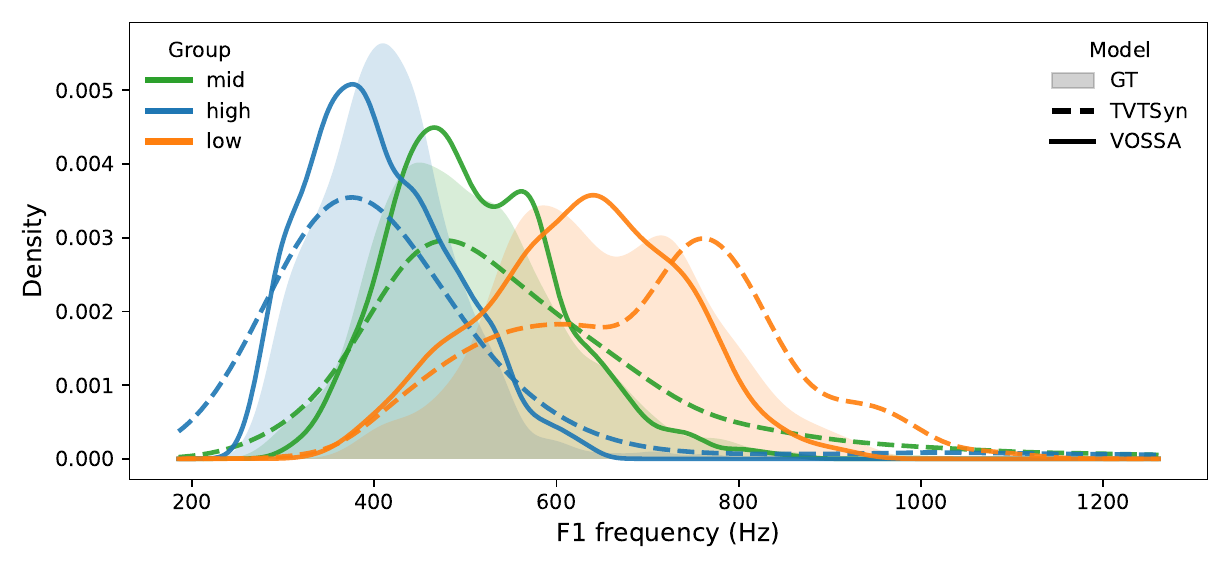}
    \caption{F1 distribution by vowel height for target speaker id00061 (Voxceleb). GT: original speech.}
    \label{fig:F1}
\end{figure}
\vspace{-5pt}
\subsection{Self-reconstruction protocol and diagnostics}
\label{sec:diagnostics}
To isolate acoustic modeling from cross-speaker effects, we also evaluate each system in a self-reconstruction setting, where the source and target speakers are identical. This protocol removes cross-speaker differences in pitch range and allows direct comparison between input and reconstructed speech. Self-reconstruction is conducted on LibriTTS.
\textbf{Pitch prediction.} We extract pitch contours using WORLD's DIO algorithm with StoneMask refinement~\cite{morise2016world, morise2009fast}. We report the pitch mean absolute error (MAE) in Hz,  the voiced/unvoiced mismatch (V/UV) ratio, and
the Pearson correlation coefficient (PCC) between predicted and ground-truth pitch trajectories (on frames where both signals are voiced). 


Shown in Table~\ref{tab:main_results}, VOSSA has the lowest pitch MAE and the highest PCC across systems, with V/UV mismatches close to those of the slt24 baseline (2nd best) and significantly lower than GenVC, which also uses an internal speaker representation. This indicates that VOSSA not only improves absolute pitch accuracy but also better preserves frame-level pitch dynamics.


\textbf{Acoustic phonetics.} As a final objective measure, we also evaluate the ability of each system to preserve formant distributions, which convey both phonetic acoustic cues and speaker-dependent cues (i.e., formant frequencies depend on vocal tract length).  For each utterance, we obtain phone-level alignments using the Montreal Forced Aligner~\cite{mcauliffe2017montreal}, extract time-aligned phoneme boundaries, and select vowel segments. We restrict the analysis to monophthongs, as diphthongs involve time-varying formant movement. Due to space constraints, we only report analysis of the first formant (F1), which is associated with vowel height. To reduce boundary and coarticulation effects, we measure F1 at the temporal midpoint of each vowel. The resulting F1 values are aggregated across utterances to form empirical distributions, which are compared across systems and ground truth to assess differences in vowel height realization.

Table~\ref{tab:main_results} reports the Wasserstein (WS) distance between resynthesized and original utterances for F1 distributions for high, mid, and low vowels \textit{on a speaker-by-speaker basis} to eliminate differences in vocal tract length across speakers. VOSSA achieves the lowest WS distance for all high, mid, and low vowels as compared to all baselines, which indicates better preservation of vowel height realizations. This result is illustrated in Fig.~\ref{fig:F1} for VOSSA against its TVTSyn backbone.


\vspace{-5pt}
\subsection{Perceptual evaluations}
To corroborate these findings, we conducted listening tests comparing VOSSA against baselines. To evaluate synthesis quality, we used a 5-point Mean Opinion Score (MOS) test (N=20 listeners; 15 utterances/model). Shown at the bottom of Table~\ref{tab:main_results}, subjective MOS ratings follow a similar trend to the NISQA-MOS results: VOSSA achieves the highest MOS, closely followed by TVTSyn and SLT24, and notably higher than GenVC-s and DarkStream. As expected, ground-truth (original) speech remains substantially higher.

For the remaining tests we compared VOSSA against its TVTSyn backbone, the overall best baseline in Table~\ref{tab:main_results}. We measured speaker similarity using an ABX test. In each trial, listeners were presented with a target utterance $X$ and samples from VOSSA and TVTSyn, $A$ and $B$, and were asked to select the sample whose speaker identity was closest to that of $X$. We also measured intelligibility and vocal vibrancy using AB preference tests. For intelligibility we asked listeners to choose the one which was easier to understand whereas for the vibrancy test they selected utterances which sounded more vibrant and/or fuller. The assignment of TVTSyn and VOSSA to $A$ and $B$ was randomized across trials to mitigate order bias. Shown in Table~\ref{tab:perception_results}, listeners prefer VOSSA across the three perceptual measures, indicating its superiority to the TVTSyn backbone.

\begin{table}[t]
\centering
\small
\setlength{\tabcolsep}{5pt}
\caption{Perceptual listening test results.}
\begin{tabular}{lcc}
\toprule
\textbf{Metric} & \textbf{TVTSyn} & \textbf{VOSSA (Ours)} \\
\midrule
Speaker similarity ($\uparrow$, \%) & 46 & \textbf{54} \\
Intelligibility ($\uparrow$, \%) & 44 & \textbf{56} \\
Vibrancy ($\uparrow$, \%) & 48 & \textbf{52} \\
\bottomrule
\end{tabular}
\label{tab:perception_results}
\end{table}
\vspace{-5pt}
\subsection{Model size and streaming efficiency}
VOSSA contains $19\%$ fewer parameters than TVTSyn (132.4M vs. 162.8M), reflecting the removal of a separate external speaker encoder. Under streaming inference (batch size 1, 16\,kHz, single NVIDIA RTX 5000 Ada GPU, 60 ms chunk size), both VOSSA and TVTSyn achieve a real-time factor (RTF) of approximately 0.25 and an end-to-end latency of about 73 ms. These results indicate that integrating speaker representation learning within the content encoder reduces overall model size without introducing additional streaming overhead. 

\section{Discussion}
We proposed VOSSA, a streaming VC system that learns speaker representations directly from intermediate content features. By eliminating a standalone ASV-based speaker encoder, VOSSA reduces model size by 19\%  while maintaining comparable streaming efficiency. Further, VOSSA attains the highest normalized target-speaker similarity and is preferred in ABX speaker similarity tests, indicating that the learned representation guides identity realization more effectively during generation. Importantly, these improvements are not limited to cosine similarity at the speaker-embedding level.  Vowel-level evaluation shows reduced Wasserstein distance to ground-truth F1 distributions across vowel heights, suggesting that VOSSA captures speaker-dependent phonetic-acoustics beyond overall speaker timbre. Finally, VOSSA achieves lower pitch error and V/UV mismatch while preserving high contour correlation, suggesting more accurate and dynamically consistent F0 modeling.

\section{Acknowledgments}
Supported by the Intelligence Advanced Research Projects Activity (IARPA) via Department of Interior/Interior Business Center (DOI/IBC) contract number 140D0424C0066. The U.S. Government is authorized to reproduce and distribute reprints for Governmental purposes notwithstanding any copyright annotation thereon. Disclaimer: The views and conclusions contained herein are those of the authors and should not be interpreted as necessarily representing the official policies or endorsements, either expressed or implied, of IARPA, DOI/IBC, or the U.S. Government.

\section{Use of Generative AI Disclosure}
LLMs were used only minimally to improve writing clarity and presentation. All experimental design, implementation, data analysis, and interpretation of results were conducted by the authors.


\bibliographystyle{IEEEtran}
\bibliography{mybib}

\end{document}